\newlength{\defaultspace}
\documentclass[11pt]{article}
\usepackage{rotating}
\usepackage{amssymb}

\begin{document}
\title { A note on "Exact and approximate methods for a one-dimensional minimax bin-packing problem" }

\author{ Mariona Vil\`a$^1$, Jordi Pereira$^2$ \\
$^1$ Escola Universit\`{a}ria d’Enginyeria T\`{e}cnica Industrial de Barcelona, \\
Universitat Polit\`{e}cnica de Catalunya. C. Comte Urgell, 187 1st. Floor, 08036 Barcelona Spain\vspace*{0.25 cm}\\
$^2$ Departamento de Ingeniería Industrial, Universidad Católica del Norte, \\
Av. Angamos 0610, Antofagasta, Chile\vspace*{0.25 cm}\\
emails: mariona.vila.bonilla@upc.edu, jorgepereira@ucn.cl}

\maketitle

\textbf{Abstract}. In a recent paper, Brusco, Köhn and Steinley [Ann. Oper. Res. 206:611-626 (2013)] conjecture that the 2 bins special case of the one-dimensional minimax bin-packing problem with bin size constraintsmight be solvable in polynomial time. In this note, we show that this problem is NP-hard for the special caseand thatit is strongly NP-hard for the general problem. We also propose a pseudo-polynomial algorithm for the special case and a constructive heuristic for the general problem.


\section{Introduction}\label{s1}
The one-dimensional minimax bin-packing problem with bin size constraints (MINIMAX\_BSC) is a special case of the bin-packing problem that appears in the area of psychology \cite{VanderLinden2005}.

The MINIMAX\_BSC case can be formally defined as follows: There are $T$ sets ($1\leq t\leq T$), and each set is composed of $B$ items ($1\leq b\leq B$). Each item has an associated weight $w_{tb}$, and these items must be grouped into $B$ groups 
in such a way that each group contains exactly one item of each one of the sets. The total weight of each group is equal to the sum of the weights of the items assigned to the group, and the objective is to minimize the maximum sum of the 
weights of the items in any group. 

In the context of test design \cite{VanderLinden2005}, the items represent questions, and each question has a level of difficulty; the sets represent groupings of questions that have comparable difficulty; the final groups represent the 
questionnaires, each of which has one question from each set. The difficulty of each questionnaire should be as even as possible, which is an objective that is achieved by minimizing the difficulty (weight) of the most difficult 
questionnaire (based on the sum of the weights). 

Brusco, K\"{o}hn and Steinley \cite{Brusco2013} recently studied the MINIMAX\_BSC, proposed a mixed zero-one integer linear programming model, and used the CPLEX commercial software to solve the model. Because this method could not 
verify optimality on large instances, they also proposed a simulated annealing (SA) algorithm to obtain near-optimal solutions. 

While \cite{Brusco2013} did not address the complexity of the problem, the analysis of the results showed that CPLEX as well as the proposed SA algorithm optimally solved large-sized instances with $B=2$, leading the authors to conjecture that this special 
case might be solvable in polynomial time.

In this note, we address the complexity of the MINIMAX\_BSC, and we propose a constructive heuristic that has an absolute performance guarantee. The remainder of this paper is structured as follows. In section \ref{s2}, we show that the 
partition  problem and the 3-partition problem can be reduced to the MINIMAX\_BSC, and we also propose a pseudo-polynomial algorithm for the case in which $B=2$. In section \ref{s3}, we present the proposed heuristic. Finally, section 
\ref{s4} provides some conclusions.

To ease the presentation of the following sections, we define the following notation: the range of the set $t$, $r_t$, is defined as $r_t=\max_{1\leq b\leq B} \{w_{tb}\} -\min_{1\leq b\leq B} \{w_{tb}\}$. The maximum range $R$ is defined as 
$R=\max_{1\leq t\leq T}\{r_t\}$. The total weight among all of the items is denoted as $W$ ($W=\sum_{1\leq t\leq T, 1\leq b\leq B} w_{tb}$). Note that a lower bound on the objective is $W/B$ (see \cite{Brusco2013}).

\section{Complexity}\label{s2}

In this section, we first reduce the partition problem to the MINIMAX\_BSC problem in which $B=2$. Furthermore, we propose a pseudo-polynomial algorithm for this case. Finally, we show that the 3-partition problem reduces to a general 
MINIMAX\_BSC problem.

\textbf{Theorem 1}. The MINIMAX\_BSC problem with $B=2$ is NP-Hard.

\textbf{Proof}.We use a reduction for the partition problem that is known to be NP-Complete \cite{Garey1979}.

\emph{Problem PARTITION}. Given a finite set $A$ and a size $s(a)\in Z^+$ for each $a\in A$, is there a subset $A’ \subseteq A$ such that $\sum_{a\in A'}s(a)=\sum_{a\in A-A'}s(a)$?

To ease the explanation of the reduction procedure, we assume that $A$ is an ordered set, and thus, we can refer to the $t$-th element of $A$ as $a_t$.

\emph{Reduction:} Given an instance of the problem PARTITION, the corresponding instance of MINIMAX\_BSC with $B=2$ is constructed as follows: Let $T$ be the cardinality of $A$ ($T=|A|$). Each set($1\leq t\leq T$) is composed of two items with 
weights $w_{t1}=a_t$ and $w_{t2}=0$. 

Clearly, the optimum objective value for this instance is equal to $\sum_{a\in A}s(a)/2$, if and only if the answer to the original instance of the Problem PARTITION is “Yes”.$\Box$

\textbf{Observation 1}. This problem can be solved in pseudo-polynomial time using a modification of the dynamic programming (DP) algorithm for the subset sum problem (see, for example, \cite{Martello1990}), which is known to be easily solvable in many 
practical applications. The DP formulation determines all of the feasible weights for one of the bins. Once the feasible weights are available, the problem consists in finding the feasible weight that minimizes the absolute difference 
between the weight and $W/2$ (which is the lower bound on the optimal solution). 

The states of the DP ($0\leq s\leq W$) identify the weights of the items that are assigned to the group. The DP has $T$ stages that are associated with the assignment of the items of set $t$ ($1\leq t\leq T$) to the group. The recurrence 
function $f_t(s)$ calculates which states are feasible (having a value equal to 1) or not (having a value equal to 0), as follows:

\begin{tabular}{ll}
	
For $t=1$,	& $f_1(w_{11})=1$, $f_1(w_{12})=1$, $f_1(s\neq w_{11} \wedge s\neq w_{12})=0$. \\

For $t=2,..,T$, & $f_2(s)= \max \{f_{t-1}(s-w_{t-1,1}); f_{t-1}(s-w_{t-1,2}) \}$. \\

\end{tabular}

The optimal solution of the problem can be obtained by reconstructing the solution from a feasible state of $f_T(s)$ with the smallest absolute value of $s-W/2$. $\Box$

\textbf{Theorem 2}.The general MINIMAX\_BSC problem is strongly NP-Hard. 

\textbf{Proof}.We use a reduction for the 3-partition problem that is known to be strongly NP-Complete [3].

\emph{Problem 3-PARTITION}. Given a set $A$ of $3\cdot m$ elements, a bound $U\in Z^+$, and a size $s(a)\in Z^+$ for each $a\in A$ such that $U/4<s(a)<U/2$ and such that $\sum_{a\in A}s(a)=m\cdot U$, can $A$ be partitioned into $m$ disjoint sets 
$A_1$,$A_2$,...,$A_m$ such that, for $1\leq i\leq m$, $\sum_{a\in A_i}s(a)=U$?

\emph{Reduction}: Given an instance of problem 3-PARTITION, the corresponding instance of MINIMAX\_BSC is constructed as follows: Let $B=m$ and $T=3\cdot m$. Then, let $w_{t1}=a_t$, $w_{tb}=0$ for $2\leq b\leq B$, $1\leq t\leq T$.

The optimum objective value for this instance of the MINIMAX\_BSC is equal to $U$ if and only if the answer to the original instance of Problem 3-PARTITION is “Yes”. $\Box$  

\section{Evaluation of the heuristic procedure }\label{s3}

We propose a fast heuristic that has $O(T\cdot B\cdot \log B)$ complexity and an absolute performance guarantee equal to $R$. Algorithm 1 depicts the algorithm\\

\textbf{Algorithm 1.}\\
Step 1. Set the accumulated weights of each group $W_b$ to $0$ for $1\leq b\leq B$.\\
Step 2. For each set $1\leq t\leq T$ do \\
Step 2a. Order the items of the set into a non-decreasing order of the weights, and order the groups into a non-increasing order of the accumulated weights. Assign the $t$-th item to the $t$-th group ($1\leq t \leq T$). \\
Step 2b. Update the accumulated weights.\\ 
Step 3. End For\\

The computationally expensive part of the algorithm is the ordering of the sets and items on the groups, which is step 2a. The for loop is executed T times, and each ordering has $O(B\cdot logB)$ complexity. The final computational 
complexity is $O(T\cdot B \cdot logB)$. 

The rationale behind the previous algorithm is to always maintain the difference among the groups within some limits. 

We now proceed to prove our claim on the performance guarantee. 

First, we show that the maximum difference of the accumulated weights between any two groups is equal to $R$. To simplify the explanation, let d denote the difference in the accumulated weights between two groups. 

During step 1, $d$ is set to 0. Accordingly, after the first assignment of items to groups, $R\geq d$ holds. During subsequent assignments, $d$ is bounded by $max\{d, r_t\}$, with $r_t$ equal to the range of the assigned set (if $r_t\leq d$,
then $d-r_t$ is bounded by $d$, and if $r_t\geq d$, then $r_t-d$ is bounded by $r_t$). Because both $d\leq R$ and $r_t\leq R$ hold, the difference between any two groups will always be smaller than $R$.

Second, note that for any valid solution, $min_{1\leq b\leq B} \{W_b\}$ is a lower bound on the optimal solution (the maximum possible value for $min_{1\leq b\leq B} \{W_b\}$ is $W/B$, which corresponds to a solution in which all of the 
groups have identical accumulated weights).  

Because the difference between any two groups obtained using Algorithm 1 is bounded by $R$, the difference between the solution provided by Algorithm 1 and a lower bound for the problem is bounded by $R$. This construct proves our initial 
claim that Algorithm 1 provides a solution that has an absolute performance guarantee equal to $R$.

\section{Conclusions}\label{s4}

In this note, we have studied the complexity of the one-dimensional minimax bin-packing problem with bin size constraints, and we have shown that the problem is NP-Hard, even if the number of bins is limited to 2. The studied problem is 
relevant in the area of psychology, as well as other areas in which the objective is to evenly distribute elements of different sets among groups.

In addition to the complexity issues, we have also proposed a constructive heuristic that has an absolute performance guarantee that is equal to the largest difference in the weights among the items on any set. While the quality of the 
solutions provided by the heuristic might not be sufficient in some circumstances, this constructive heuristic is very fast. 

We implemented Algorithm 1 and ran some limited tests on instances that were generated as proposed in \cite{Brusco2013} and \cite{VanderLinden2005}. Our results indicate that (1) the experimental efficiency of Algorithm 1 is improved if 
the sets are ordered according to non-increasing values of $r_t$; (2) the running times for the instances that have up to 6000 items and 300 groups are below 0.02 seconds on a current commodity computer (3.06 GHz Intel Core 2 Duo); and (3) 
the average gap between the solution provided by the heuristic and the trivial lower bound is below 0.07.

While the SA procedure from \cite{Brusco2013} should be the recommended solution method for this problem if the running time is not a limiting constraint, these results indicate that the SA could be improved by initializing the search 
using the proposed algorithm rather than randomly generating the initial solution. We believe that this modification would reduce the running time that is required by the SA algorithm to reach good solutions.


\end{document}